\documentclass[letterpaper,12pt]{article}   
\usepackage{osajnl2} 
\usepackage[draft]{hyperref} 

\begin{document}

\title{Polarization manipulation holographic lithography by single refracting prism}


\author{Yi Xu, Man Wu, Xiuming Lan, Xiaoxu Lu,Sheng Lan and Lijun Wu$^{*}$}
\address{Laboratory of Photonic Information Technology, School for Information and Optoelectronic Science and Engineering,\\ South China Normal University, Guangzhou 510006, P.R. China\\}

\address{$^*$Corresponding author: ljwu@scnu.edu.cn}

\begin{abstract}We propose a simple but effective strategy for polarization manipulation holographic lithography by single refractive prism. By tuning the polarization of single laser beam, we simply obtain a pill shape interference pattern where multiple modulated beams are needed in conventional holography lithography. Fabrication of large area pill shape two-dimensional polymer photonic crystals template using one beam and one shoot holography lithography is shown as an example. This integrated fabrication technique, for example, can release the crucial stability restrictions imposed on the multiple beams holography lithography.
\end{abstract}

\ocis{230.5480,230.5440,090.2890,090.2880,090.4220}

\maketitle 

\section{Introduction}
Holographic lithography has been demonstrated as an effective and inexpensive way to fabricate Photonic crystals(PhCs) circuit~\cite{nature,labchip}. Combining with the interference between several noncoplanar beams, all five 2-D and all fourteen 3-D Bravais lattices can be obtained with controllable shape of the unit cell~\cite{14,shape}. For conventional holographic lithography combing with multiple beams, a laser is split into several beams with different optical paths and then all the beams are superimposed at the exposure region. This kind of realization needs crucial alignment accuracy which is difficult to guarantee during the whole process. Adjusting the polarization state and phase of each individual beam offers extra degrees of freedom to tailor the atom shape of the unit cell. However, several wave plates or phase delay elements should be inserted to the corresponding optical paths which brings extra stability requirement of these complex experiment setup~\cite{wang1,zhong1,liang}. All these optical elements make the coherent system very sensitive to any vibrational instability in the environment that indicates the conventional holographic lithography isn't the best way for large volume PhCs fabrication or other processes using holography~\cite{trapping}.

\begin{figure}[htbp]%
\centering
\includegraphics*[width=8cm,height=7cm]{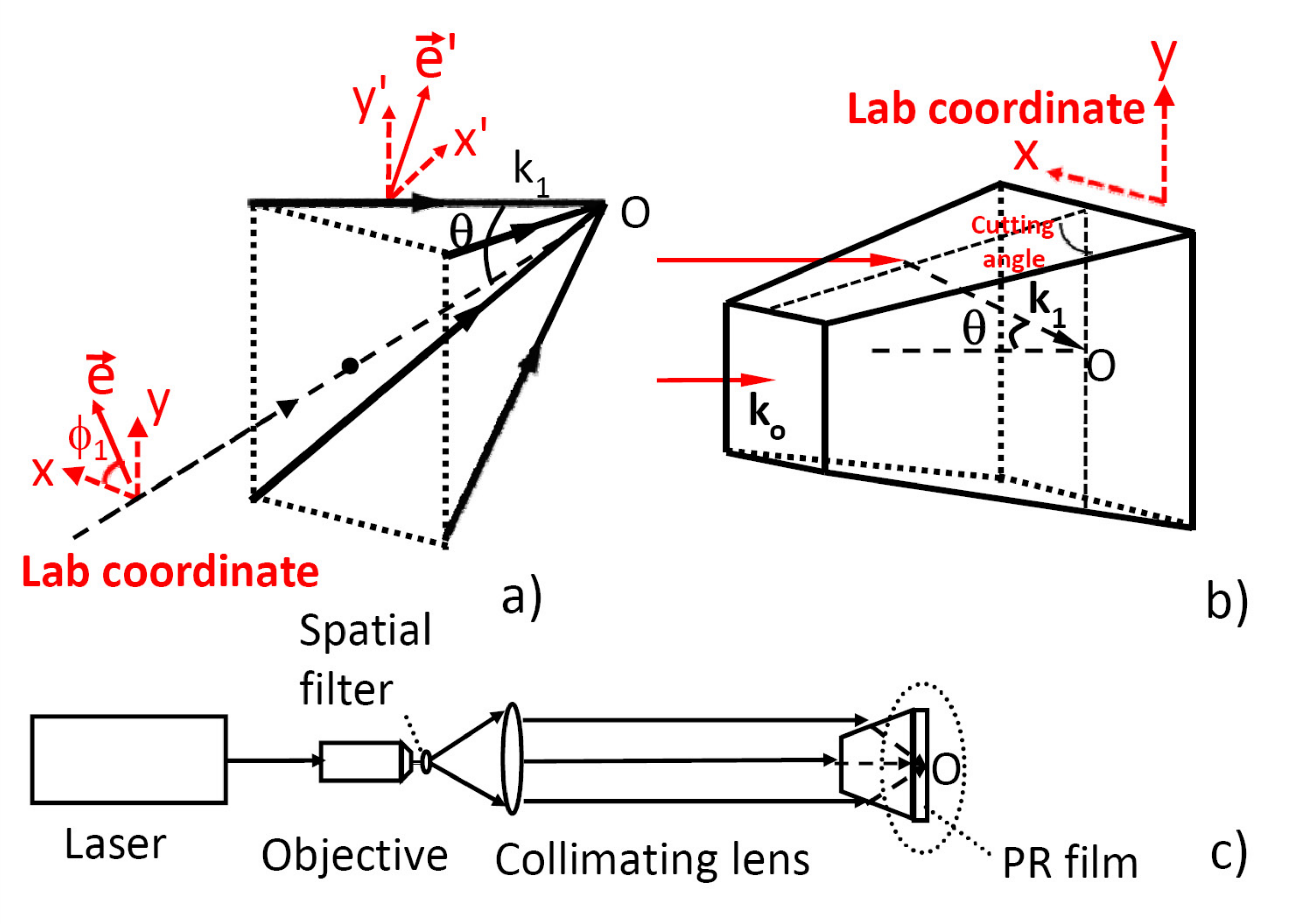}
\caption{%
 (a) A typical four beams geometry. The coordinates are outlined by dashed red arrows($ x $ and $ y $ are the lab coordinates while $ x' $ and $ y' $ are the rotated beam coordinates) . (b) The specially designed refracting prism with cutting angle $ 62^{o} $. (c) Sketch map of the optical setup used for the fabrication of PhCs. }
\end{figure}

Diffraction element mask can be used to obtained 3D PhCs in photosensitive polymer, by which the alignment and stability of the fabrication setup can be improved~\cite{stable,ao1}. Alternatively, by using an expanded laser and a single refractive prism, the feasibility of fabricating large area two- and three-dimensional defect free polymer PhCs was demonstrated~\cite{wu1}. Such simple while stable set-up can be generalized to fabricate more sophisticated periodic structure~\cite{yang1,zhong2,tam,wu2,1550,ao2,sg,glass}. Adding optical elements into such simple process facilitates the tunable production~\cite{zhong2,glass}. Very recently, by using a prism with different lateral dimensions(adjusting the phase delay between different beam path), S. G. Park et. al. have fabricated a high quality woodpile PhCs template for $ Cu_{2}O $ inversion~\cite{advance}. However, manipulating the polarizations of the interfering beams by the prism itself haven't been addressed before.

In this paper, we show that such prism naturally integrates the function of polarization manipulation which would further release the stable requirement of the holography lithography process. As a particular example, we demonstrate such integrated setup can be used to fabricate template for PhCs with polarization-independent self-collimation characteristic~\cite{yi}.

\section{Theoretical design}
The interference pattern formed by N noncoplanar beams(indicated by Fig. 1(a)) can be expressed as:
\begin{equation}
I=Re(\sum_{l,m=1}^{N} \textbf{$ {e_{l}}^{*} $}\cdot \textbf{$ {e_{m}} $}\exp(i(k_{l}-k_{m})\cdot \textbf{r}+i(\delta_{l}-\delta_{m})))
\end{equation}
where $ e $ stands for the unit electric vector, $ k $ represents the k-vector of each beam and $ \delta $ is phase delay. The k-vectors of the beams determine the translation and rotation symmetry while the intensities and the polarization states of the beams have an effect on the atom shape in the unit cell of the PhCs. Considering a linear polarized light imprints on a dielectric interface, it suffers from reflection and refraction. The polarization component laying at the direction of the incident plane normal(S wave) would not change its direction while the component laying at the incident plane(P wave) would change its direction because of the electromagnetic field's boundary condition. Therefore, different cutting faces of the prism thus obtain different refraction angles and polarization directions referring to the lab coordinates as is indicated in Fig. (1) (a). It means that the prism not only integrates the function of beam splitter and combiner but also the function of polarization management. Without loss of general, we choose four beams arranging symmetrically as an example and the cutting angle of the prism is $ 62^{o} $[shown in Fig. (1) (b)]. We plot output polarizations(the electric field component of refractive beam 1 at the Lab coordinates) in Fig. 2 (a) when the input polarization($ \phi_{1} $) varying from $ 0^{o} to $ $ 360^{o} $. It can be seen from the figure that the polarization direction of the refracted beam can be controlled by simply adjusting the input polarization. All directions inside the plane normal to the k-vector of the refractive beam can be obtained. At the same time, the output electric vectors of the beams refracting from different cutting faces are distinct with certain input polarization and can be tuned via the input beam[Fig. 2 (b)]. Please note that there are double degenerate output polarization dependences because of the 4 folds rotation symmetry of the prism in these case.  

\begin{figure}[t]%
\centering
\includegraphics*[width=5.5cm,height=8cm]{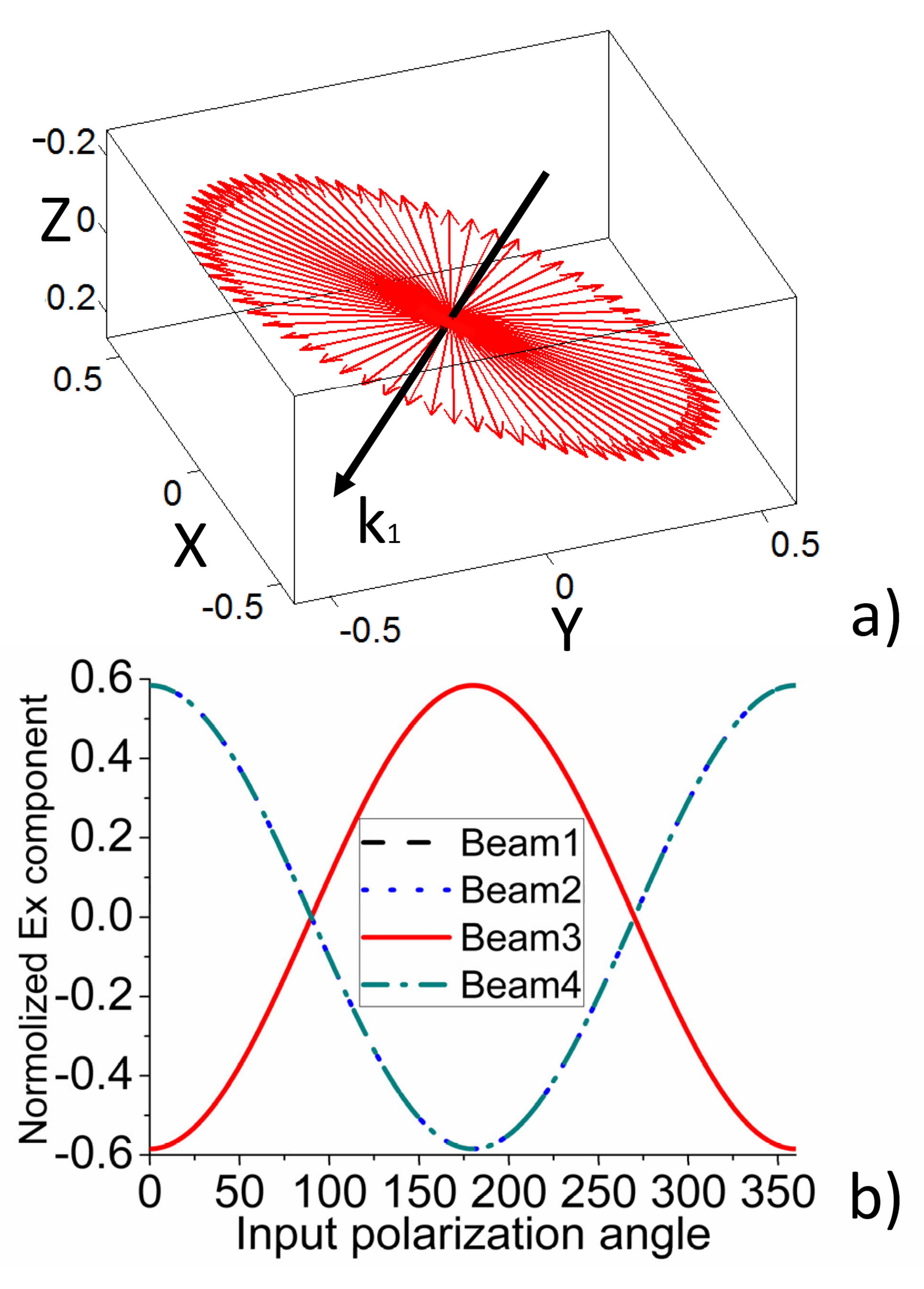}
\caption{%
 (a)All output polarizations vectors(polarization component of beam 1) when the input polarization (rotating from $ 0^{o} $ to $ 360^{o} $). (b) The dependence of output polarizations(x-axis component of beam 1-4) on the input polarization.}
\end{figure}

Pill shape PhCs have the advantage to minimize the spatial dispersion discrepancy between the transverse electric wave and transverse magnetic wave in the first photonic band~\cite{yi}. Fabricating such PhCs by conventional lithography~\cite{shape,wang1,zhong1,liang} needs several wave plates to manage the polarization state and definitely impose restrictions on the stability of the interference process. We thus look into this example to produce PhCs by our simple setup. It should be pointed out that such method is not limit to this specified example while it can be generalized to fabricate PhCs with different rotation symmetry. 

In principle, the PhCs with square symmetry acquire the interfering beams arranging in specified symmetry which produces the necessary reciprocal lattice. We thus choose three beams arranging with the azimuth angles $ 0^{o},180^{o},270^{o} $. The computer optimized input polarized direction for producing the pill shape PhCs is $ \phi_{1}=90^{o} $ at which the interference result generates the target pill shape pattern and has the highest contrast ratio. Now let's look into the experiment demonstration which outlines the feasibility of this scheme. 

\section{Experiment procedure}
The experiment sketch map is shown by Fig. 1 (c). A continuous-wave laser at 532 nm ($Verdi 5,Coherent$) shines on the chopped-off prism, after which passed the spatial filtering and expand system. The top of the prism and one of the cutting facet are blocked to obtain three beams with desired azimuth angles. The expanded beam would be split and the interference region is at the bottom of the prism. Fig. 3 (a) and (b) show the numerical result and the patten recorded by a CCD, which confirms our proposal. During the exposure process, an angled thick glass window was attached to the substrate to reduce the effects of back-reflected laser light. All the interfaces are filled with Silicone oil to avoid the total-reflection. The chemical treatment is the same as ~\cite{wu1}. As can be seen from the SEM pictures in Fig. 3 (c), the PhCs with pill shape atoms arranging in square symmetry are achieved. While in Fig. 3(d), large area pill shape PhCs confirm the high throughput capacity of such simple polarization manipulation holographic lithography. The size of the pill shape atom can be controlled by the exposure time.


\begin{figure}[t]%
\centering
\includegraphics*[width=9cm,height=9cm]{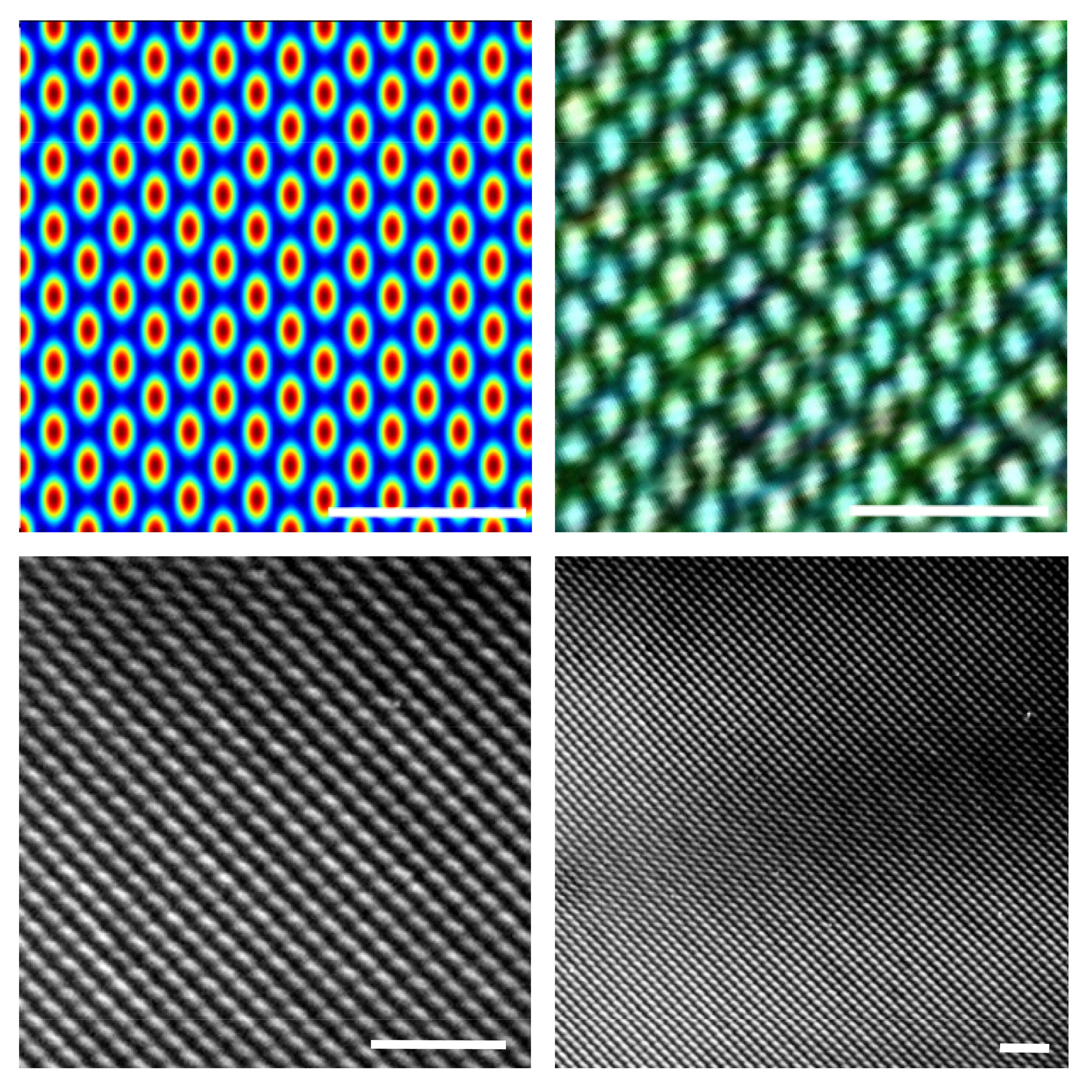}
\caption{%
  (a) Numerical calculation of the interference at the bottom of the prism when a polarized expanded beam($ \phi_{1}=0 $) shines on it. (b) Interference pattern recorded by a CCD. (c) Zoom-in view of the fabricated PhCs. (d)Bird's-eye view of the fabricated PhCs. All white bars indicate the same length scale of 5$ \mu $ m }
\end{figure}

\section{Conclusion}
To summarize, we propose theoretically and demonstrate experientially a compact method of polarization control in the holography lithography which, for example, can be used to produce PhCs template. The electric vector of each interfering beam can be simply manipulated via the input expanded beam which make the whole process more stable. Pill shape PhCs template as a particular example is fabricated in the polymer. We expect this technique can find application in other holography system~\cite{trapping}.

\end{document}